\documentclass[twocolumn,epjc3]{svjour3}

\usepackage{amsmath,amssymb}

\usepackage{url}
\usepackage{bm}
\usepackage{graphicx}
\usepackage{epsfig}
\usepackage{rotating}
\usepackage{dcolumn}
\usepackage{xcolor}
\usepackage{slashed}
\usepackage{threeparttable}
\usepackage[colorlinks=true
,urlcolor=blue
,anchorcolor=blue
,citecolor=blue
,filecolor=blue
,linkcolor=blue
,menucolor=blue
,pagecolor=blue
,linktocpage=true
,pdfproducer=medialab
,pdfa=true
]{hyperref}

\RequirePackage{mathptmx}
\RequirePackage{latexsym}
\RequirePackage[numbers,sort&compress]{natbib}

\newcommand{\beq}{\begin{eqnarray}} 
\newcommand{\eeq}{\end{eqnarray}} 
\journalname{Eur. Phys. J. C}

\begin{document}

\title{
\boldmath 
$W^+_{} W^-_{} H$ production at lepton colliders:
A new hope for heavy neutral leptons
}


\author{J. Baglio\thanksref{e1,addr1,addr2,addr3}
  \and
  S. Pascoli\thanksref{e2,addr3}
  \and
  C. Weiland\thanksref{e3,addr3}}
\thankstext{e1}{email: \href{mailto:julien.baglio@uni-tuebingen.de}{julien.baglio@uni-tuebingen.de}}
\thankstext{e2}{email: \href{mailto:silvia.pascoli@durham.ac.uk}{silvia.pascoli@durham.ac.uk}}
\thankstext{e3}{email: \href{mailto:cedric.weiland@durham.ac.uk}{cedric.weiland@durham.ac.uk}}


\institute{
  Institut f\"{u}r Theoretische Physik, Eberhard Karls
  Universit\"{a}t T\"{u}bingen, Auf der Morgenstelle 14, D-72076
  T\"{u}bingen, Germany\label{addr1}
  \and
  Institute for Advanced Study, Durham University, Cosin's
  Hall, Palace Green, Durham DH1 3RL, United~Kingdom\label{addr2}
  \and
  Institute for Particle Physics Phenomenology, Department
  of Physics, Durham University, South Road, Durham DH1 3LE,
  United~Kingdom\label{addr3}
}

\date{October 4, 2018}

\maketitle

\begin{abstract} 
We present the first study of the production of a Standard Model Higgs
boson at a lepton collider in association with a pair of $W$ bosons,
$e^+_{} e^-_{} \to W^+_{} W^-_{} H$, in the inverse seesaw
model. Taking into account all relevant experimental and theoretical
constraints, we find sizable effects due to the additional heavy
neutrinos up to $-38\%$ on the total cross-section at a center-of-mass
energy of 3~TeV, and even up to $-66\%$ with suitable cuts. This
motivates a detailed sensitivity analysis of the process $e^+_{}
e^-_{} \to W^+_{} W^-_{} H$ as it could provide a new, very
competitive experimental probe of low-scale neutrino mass models.
\end{abstract}


\section{Introduction}

Neutrino oscillations, as discovered by the Super-Kamiokan\-de
experiment in 1998~\cite{Fukuda:1998mi} and subsequently confirmed by
a plethora of results~\cite{Patrignani:2016xqp}, imply that at least
two neutrinos have a non-zero mass. This cannot be explained in the
Standard Model (SM) and calls for an extension of this framework. One
of the simplest possibilities is the addition of new fermionic
gauge-singlet states that play the role of right-handed neutrinos,
leading to the type-I seesaw mechanism and its
variants~\cite{Minkowski:1977sc,Ramond:1979py,GellMann:1980vs,Yanagida:1979as,Mohapatra:1979ia,Schechter:1980gr,Schechter:1981cv,Mohapatra:1986aw,Mohapatra:1986bd,Bernabeu:1987gr,Pilaftsis:1991ug,Ilakovac:1994kj,Akhmedov:1995ip,Akhmedov:1995vm,Barr:2003nn,Malinsky:2005bi}. Amongst
the various seesaw realizations, one of particular interest is the
inverse seesaw model
(ISS) \cite{Mohapatra:1986aw,Mohapatra:1986bd,Bernabeu:1987gr}. It was
proved in~\cite{Moffat:2017feq} that, in any model that only adds
fermionic gauge singlets to the SM field content with no cancellation
between the contributions to the light neutrinos masses from different
orders of the seesaw expansion or different radiative orders,
requiring the light neutrinos to be massless is equivalent to
requiring lepton number to be conserved. The inverse seesaw verifies
all these conditions and we indeed observe that in the
lepton-number-conserving limit of this model, light neutrinos are
massless, independently from the seesaw scale or the size of the
neutrino Yukawa coupling. In this renormalizable, testable, low-scale
seesaw model, light neutrino masses are suppressed not by a
small-active sterile mixing as in the high-scale type I
seesaw. Instead, this model relies on an approximately conserved
lepton number in agreement with the theorem~\cite{Moffat:2017feq},
thus allowing to generate the light neutrino masses while having large
neutrino Yukawa couplings and heavy sterile neutrinos at the TeV
scale, opening the exciting possibility of detecting the latter in
current or future planned high-energy colliders, see for example
Refs.~\cite{Deppisch:2015qwa,Antusch:2016ejd,Cai:2017mow} for
reviews. It is particularly worth noting that this model provides a
prototype of fermionic low-scale seesaw, making our results applicable
to a wide range of models.

As the neutrino Yukawa couplings in the ISS can be large, the 
properties of the Higgs boson, the remnant of the electroweak
symmetry-breaking
mechanism~\cite{Higgs:1964ia,Englert:1964et,Higgs:1964pj,Guralnik:1964eu}
generating the masses of the other fundamental particles in the SM
and that was discovered at the Large Hadron Collider (LHC) in
2012~\cite{Aad:2012tfa,Chatrchyan:2012ufa}, can be sizeably
affected. This opens new search strategies which rely on the Higgs
boson, for instance Higgs
decays~\cite{BhupalDev:2012zg,Cely:2012bz,Gago:2015vma,Abada:2018sfh},
searches in Higgs production at lepton
colliders~\cite{Antusch:2015mia,Antusch:2015gjw}, or lepton flavour
violating Higgs decays~\cite{Arganda:2004bz,Arganda:2014dta}. We also
investigated recently the heavy neutrino impact on the triple Higgs
coupling~\cite{Baglio:2016ijw,Baglio:2016bop}.

Based on the idea that $t$--channel fermions coupled to a Higgs boson
can give sizeable contributions to a cross-section, see for example the
case of $b\bar{b}\to W^+_{} W^-_{} H$ at the
LHC~\cite{Baglio:2016ofi}, we investigate in this paper, for the first
time, the impact of heavy neutrinos on the production of a Higgs boson
in association with a pair of $W$ bosons at a lepton collider, $e^+_{}
e^-_{} \to W^+_{} W^-_{} H$. This process has been studied in the SM
and has been found to have good detection
prospects \cite{Baillargeon:1993iw}. We describe the ISS model and
discuss the relevant theoretical and experimental constraints. We
present our calculational setup before a numerical analysis of our
results is carried out. Performing a scan over the relevant parameters
of the model, we find deviations up to $-38\%$ on the total
cross-section at 3~TeV, that can be enhanced to $-66\%$ after applying
a reasonable set of cuts that leaves an ISS cross-section of
0.14~fb. We also provide a simplified formula which reproduces our
results within one percent.

\section{The model and its constraints}

The ISS model~\cite{Mohapatra:1986aw,Mohapatra:1986bd,Bernabeu:1987gr}
is an appealing low-scale seesaw model  that extends the SM with
fermionic gauge singlets. We consider here a realisation where each
generation is supplemented with a pair of these right-handed gauge
singlets, $\nu_R^{}$ and $X$, which have opposite lepton number. This
provides a realistic realisation of seesaw models close to the
electroweak scale that can reproduce low-energy neutrino masses and
mixing while being in agreement with all experimental bounds. The
additional mass terms to the SM Lagrangian are
\begin{equation}
  \label{LagrangianISS}
  \mathcal{L}_\mathrm{ISS} = - Y^{ij}_\nu \overline{L_{i}}
  \widetilde{\Phi} \nu_{Rj} - M_R^{ij} \overline{\nu_{Ri}^C} X_{Rj} -
  \frac{1}{2} \mu_{X}^{ij} \overline{X_{i}^C} X_{j} + \mathrm{ h.c.}\,,
\end{equation}
where $\Phi$ is the SM Higgs field and $\widetilde \Phi=\imath
\sigma_2 \Phi^*$, $i,j=1\dots 3$, $Y_\nu$ and $M_R$ are complex
matrices and $\mu_{X}$ is a complex symmetric matrix.
A major characteristic of the ISS is the presence of a naturally small
lepton-number-breaking parameter $\mu_{X}$ to which the light neutrino
masses are proportional. Indeed after block-diagonalising the full
neutrino mass matrix, the $3\times 3$ light neutrino mass matrix is
given by~\cite{GonzalezGarcia:1988rw}
\begin{equation}
 \label{MlightLO}
 M_{\mathrm{light}} \simeq m_D M_R^{T-1} \mu_X M_R^{-1} m_D^T\,,
\end{equation}
at leading order in the seesaw expansion parameter $m_D M_R^{-1}$,
where $m_D=Y_\nu \langle \Phi\rangle$. This decouples the light
neutrino mass generation from the mixing between active and sterile
neutrinos (that is proportional to $m_D M_R^{-1}$) and allows for
large Yukawa couplings even when the seesaw scale is close to the
electroweak scale. It is worth noting that in this model, the heavy
neutrinos form pseudo-Dirac pairs where the splitting is controlled by
$\mu_X$ as can be seen from diagonalizing the 1-generation neutrino
mass matrix, which gives
\begin{equation}
 m_{N_1,N_2}=\pm \sqrt{M_R^2+m_D^2} + \frac{M_R^2 \mu_X}{2(M_R^2+m_D^2)}\,,
\end{equation}
in the seesaw limit $\mu_X \ll m_D, M_R$~\cite{Arganda:2014dta}.

Since one of the main motivations of our model is to explain neutrino
oscillations, we reproduce low-energy data from the global fit {NuFIT}
3.0~\cite{Esteban:2016qun} by using\linebreak the $\mu_X$-parameterisation
adapted to include next-order terms in the seesaw expansion that are
relevant for large active-sterile mixing~\cite{Baglio:2016bop}
\begin{align}
  \label{muXparam}
  \begin{split}
    \mu_X \simeq
    & \left(\mathbf{1}-\frac{1}{2} M_R^{*-1} m_D^\dagger m_D M_R^{T-1}
    \right)^{-1}\, \times \nonumber \\
    & M_R^T m_D^{-1} U_{\rm PMNS}^* m_\nu U_{\rm
      PMNS}^\dagger m_D^{T-1} M_R\, \times\\
    & \left(\mathbf{1}-\frac{1}{2} M_R^{-1} m_D^T m_D^*
      M_R^{\dagger-1}\right)^{-1}\,.
  \end{split}
\end{align}
$m_\nu$ is the diagonal light neutrino mass matrix and $U_{\rm PMNS}$
is the unitary Pontecorvo-Maki-Nakagawa-Sakata
(PMNS)~\cite{Pontecorvo:1957cp,Maki:1962mu} that diagonalises
$M_{\mathrm{light}}$. We have chosen $\delta=0$ for the Dirac CP phase
in $U_{\rm PMNS}$ for simplicity. We fix the lightest neutrino mass to
$0.01$~eV, in agreement with the Planck
results~\cite{Ade:2015xua}. The strongest experimental constraints for
this study come from a global fit~\cite{Fernandez-Martinez:2016lgt} to
electroweak precision observables (EWPO), tests of CKM unitarity and
tests of lepton universality. Since we choose all mass matrices and
couplings in the neutrino sector to be real and consider diagonal
Yukawa couplings $Y_\nu$ in our study, we do not expect electric
dipole moment measurements and lepton-flavour-violating processes to
provide relevant constraints in this scenario. Finally, we require
that the Yukawa couplings $Y_\nu$ remain perturbative, namely
\begin{equation}
 \frac{|Y_{ij}|^2}{4 \pi} < 1.5\,.
\end{equation}

\section{Calculational details}

The cross-section is calculated at leading order (LO), both in the SM
and in the ISS. Next-to-leading order electroweak corrections have
been calculated in the SM~\cite{Mao:2008qy} and are found to be
negligible for center-of-mass (c.m.) energies above 600 GeV and of the
order of $-2\%$ at $\sqrt{s}=500$~GeV, that would correspond to the
lowest International Linear Collider c.m. energy that would be
relevant for our process~\cite{Baer:2013cma}. Given the size of the
ISS deviation we obtain (of the order of $-5\%$ at $\sqrt{s}=500$~GeV
and down to $-38\%$ at higher c.m. energies, see later), we will not
take these electroweak corrections into account in our analysis.

\begin{figure}[!h]
  \centering
  \includegraphics[scale=0.53]{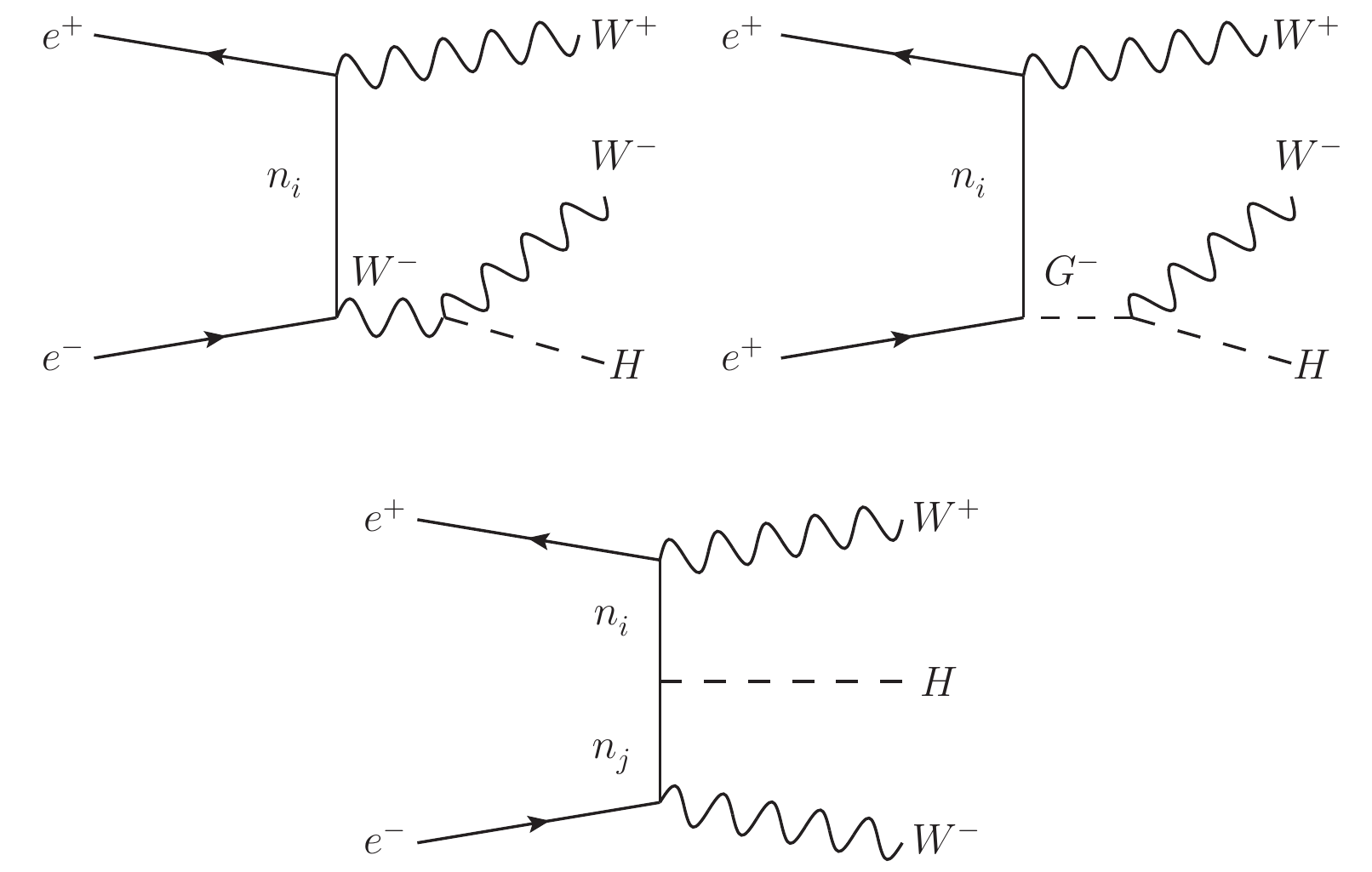}
  \caption[]{Generic Feynman diagrams representing the ISS neutrino
    contributions to $e^+_{}e^-_{}\to W^+_{} W^-_{} H$ in the
    Feynman-'t~Hooft gauge. The indices $i,j$ run from 1 to 9. Mirror
    diagrams are obtained by flipping all the electric charges.}
  \label{fig:feyndiags}
\end{figure}

The charged leptons are taken massless and their coupling to the Higgs
boson is neglected. The calculation is done in the Feynman-'t~Hooft
gauge. The Feynman diagrams at LO include $s$--channel exchanges of a
$Z$ boson or a photon, as well as $t$--channel diagrams involving the
neutrinos for which a generic selection is displayed in
fig.~\ref{fig:feyndiags}. The remaining $t$--channel diagrams are
obtained with a flipping of the $W$ and charged Goldstone boson
contributions from the $W^-_{}$ line to the $W^+_{}$ line. We have
used our own ISS model file developed for the packages {\tt FeynArts
  3.7} and {\tt FormCalc 7.5}~\cite{Hahn:1998yk,Hahn:2000kx} to
generate a Fortran code, and the numerical integration has been
performed with {\tt BASES 5.1}~\cite{Kawabata:1995th} in order to
obtain a selection of kinematic distributions.

Similar to the SM calculation, the interference terms are significant
and destructive. The dominant contribution to the ISS amplitude comes
from the first two diagrams in fig.~\ref{fig:feyndiags} with heavy
neutrinos and which go as\linebreak  $|Y_\nu|^2 v^2/M_R^2 (a+b v^2/M_R^2)$, and
from the third diagrams with one heavy neutrino and one light neutrino
in the $t$--channel which goes as $|Y_\nu|^2 v^2/M_R^2$, in
terms of the seesaw parameters. In order to enhance the cross-section
we have also performed a calculation with polarised beams. More
specifically, we have chosen, based on the Compact Linear Collider
(CLIC) baseline~\cite{CLIC:2016zwp}, an unpolarised positron beam,
$P_{e^+_{}}^{} = 0$, and a polarised electron beam with
$P_{e^-_{}}^{}=-80\%$. If we define $\sigma_{LR(RL)}^{}$ as the
cross-section for a completely polarised left-handed (right-handed)
positron with $P_{e^+_{}}^{} = -100\% \linebreak (+100\%)$ and a completely
polarised right-handed\linebreak (left-handed) electron with
$P_{e^+_{}}^{}=+100\%(-100\%)$, the polarised cross-section for
arbitrary polarisation fractions $P_{e^+_{}/e^-_{}}^{}$ can be written
as~\cite{MoortgatPick:2005cw}
\begin{align}
  \sigma_{\rm pol}^{} = \frac14\Big[ & (1-P_{e^+_{}}^{})(1+P_{e^-_{}}^{})
  \sigma_{LR}^{} + (1+P_{e^+_{}}^{})\\
  {} & \left. (1-P_{e^-_{}}^{})\sigma_{RL}^{}\right],\nonumber
\end{align}
since the LL and RR cross-sections are identically zero in our
process.

\section{Numerical results}

The calculation is done in the $G_\mu^{}$ scheme (see
e.g. Ref.~\cite{Denner:2003iy}, and Ref.~\cite{Antusch:2006vwa} in the
context of neutrino mass models) and the input parameters are the $Z$
mass $M_Z^{}$, the $W$ boson mass $M_W^{}$ and the Fermi constant
$G_F^{}$. Including the Higgs mass $M_H^{}$, the parameter values are
chosen as
\begin{align}
  M_W^{} = 80.385~\text{GeV}\,,\,\, M_Z^{} =
  91.1876~\text{GeV}\,,\nonumber
\end{align}
\vspace{-9mm}
\begin{align}
  M_H^{} = 125~\text{GeV}\,,\,\,  G_F^{} = 1.16637\times
  10^{-5}_{}~\text{GeV}^{-2}_{}\,.
\end{align}
Based on our previous analysis on the triple Higgs
coupling \cite{Baglio:2016bop}, we use the $\mu_X^{}$-parameterisation
with a degenerate Yukawa texture, $Y_\nu^{} = |Y_\nu^{}| I_3^{}$, with
hierarchical heavy neutrinos, $M_R^{} = \text{diag}(M_{R_1^{}}^{},
M_{R_2^{}}^{},M_{R_3^{}}^{})$. To illustrate our results we select the
same hierarchy as in~\cite{Baglio:2016bop},
\begin{align}
M_{R_1^{}}^{} = 1.51 M_R^{},\,\, M_{R_2^{}}^{} = 3.59 M_R^{},\,\,
  M_{R_3^{}}^{} = M_R^{}.
\label{eq:hierarchical}
\end{align}
From now on, $M_R^{}$ is to be understood as a number as well in a
slight abuse of notation. These specific heavy neutrino mass ratios
are related to our choice of $Y_\nu=|Y_\nu^{}| I_3^{}$ since they make
the constraints of the global fit~\cite{Fernandez-Martinez:2016lgt}
impact every generation similarly.

\begin{figure}[!h]
  \centering
  \includegraphics[scale=0.75]{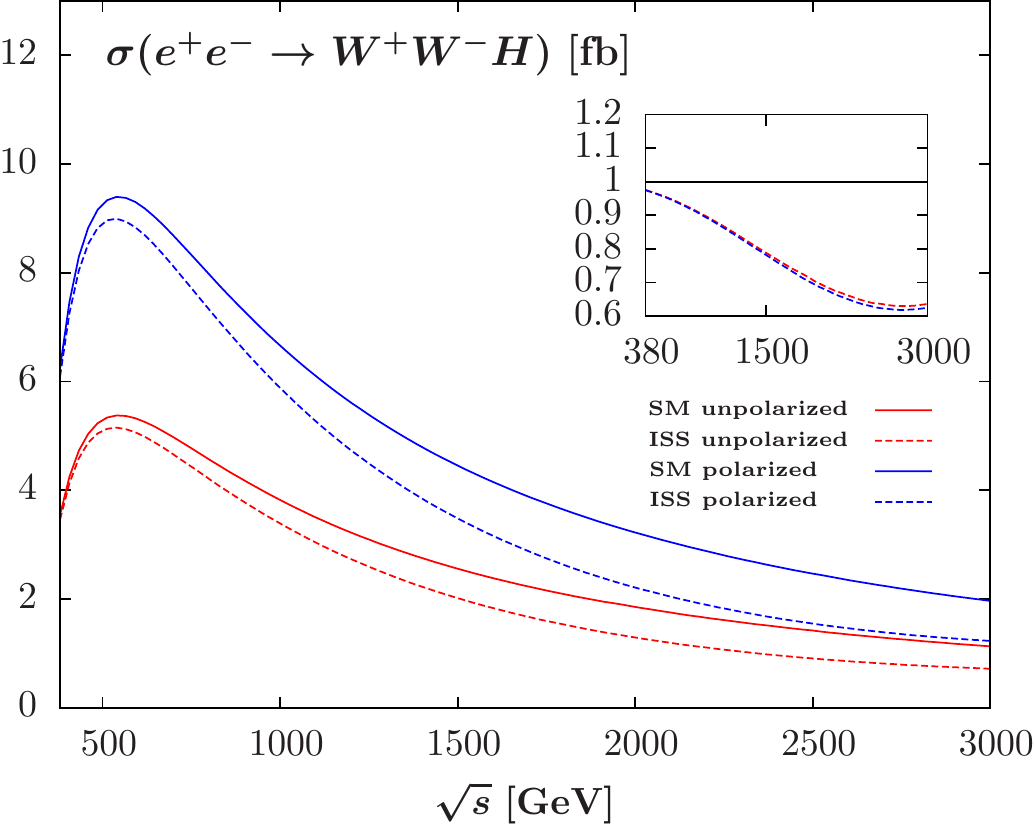}
  \caption[]{LO total $W^+_{}W^-_{} H$ production cross-section at an
    electron-positron collider (in fb) as a function of the
    c.m.~energy $\sqrt{s}$ (in GeV). The solid curves stand for the SM
    predictions, the dashed curves stand for the ISS predictions using
    the benchmark scenario described in the text. The red (blue)
    curves are for an unpolarised ($-80\%$ polarised electron beam)
    cross-section. The ratio of the ISS cross-section with respect to
    the SM prediction is shown in the insert.}
    \label{fig:xsbenchmarkf1}
\end{figure}

We present in fig.~\ref{fig:xsbenchmarkf1} the variation of the total
production cross-section $\sigma(e^+_{}e^-_{}\to W^+_{}W^-_{}H)$ as a
function of the c.m.~energy $\sqrt{s}$, using a benchmark scenario
with $|Y_\nu^{}| = 1$ and $M_R^{} = 2.4$~TeV, resulting in a
heavy neutrino spectrum with three pairs of pseudo-Dirac neutrinos of
mass 2.4~TeV, 3.6~TeV, and 8.6~TeV. We stress that this scenario, with
reasonable $\mathcal{O}(1)$ Yukawa couplings, is allowed by current
experimental and theoretical constraints.

The gain by going from an unpolarised cross-section to the polarised
electron beam is illustrated by the factor-of-two difference between
the red curves (unpolarised) and the blue curves (polarised). The
behaviour of the ISS contribution in the polarized cross-section is
the same as that of the unpolarized one, meaning that the use of a
polarized beam will lead to more events thus increasing the
sensitivity to the large deviations coming from the ISS.
The maximum of the cross-section is obtained at c.m.~energies
around 500 GeV, for which the ratio of the ISS cross-section with
respect to the SM cross-section, shown in the insert, is around
0.95. The negative contribution from the ISS correction increases
with higher c.m.~energies, reaching already $20\%$ for $\sqrt{s}\sim
1.4$~TeV and a maximal deviation of $-38\%$ at a c.m.~energy close to 3
TeV, from which the ISS correction starts to decrease for increased
c.m.~energies.

\begin{figure}[!h]
  \centering
  \includegraphics[scale=0.72]{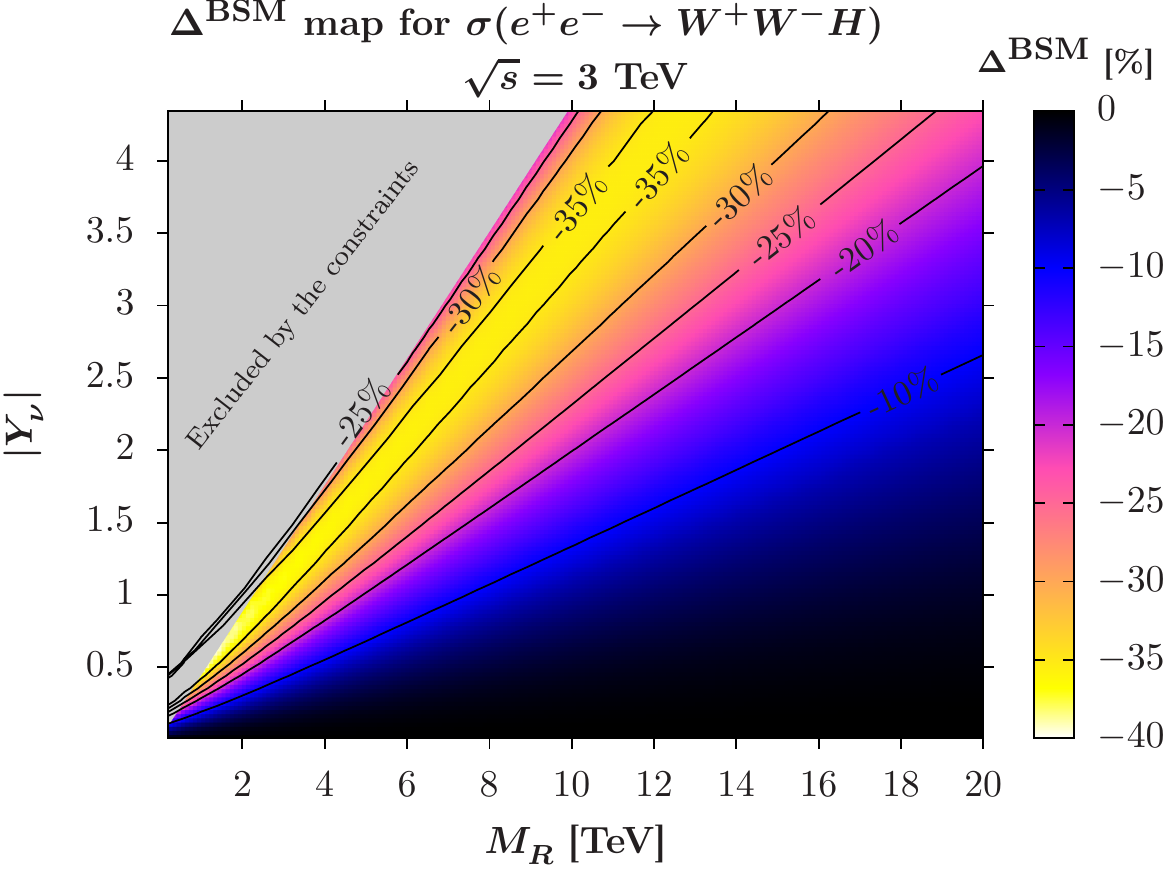}
  \caption[]{Contour map of the neutrino corrections $\Delta_{}^{\rm
       BSM}$ (in percent) to the $W^+_{}W^-_{} H$ production
     cross-section at a 3 TeV electron-positron collider,
     using a $-80\%$ polarised electron beam, as a function of the
     seesaw scale $M_R^{}$ (in GeV) and $|Y_\nu^{}|$ in the
     $\mu_X^{}$-parameterisation, using a diagonal Yukawa texture and a
     hierarchical heavy neutrino mass matrix with the parameters
     defined in eq.(\ref{eq:hierarchical}). The grey area is excluded
     by the constraints.}
   \label{fig:xsscan}
\end{figure}

In order to get insights into the dependence of the ISS correction on
the parameters of the ISS, we have performed in fig.~\ref{fig:xsscan}
a scan of the ISS deviation with respect to the SM production
cross-section, $\Delta^{\rm BSM}_{} = (\sigma^{\rm ISS}_{}-\sigma^{\rm
  SM}_{})/\sigma^{\rm SM}_{}$, as a function of the seesaw
scale $M_R^{}$ and of the parameter $|Y_\nu^{}|$ for the diagonal
Yukawa texture we have chosen and still using heavy hierarchical
neutrinos with the parameters of eq.(\ref{eq:hierarchical}). The
c.m.~energy is fixed to $\sqrt{s} = 3$~TeV which is the last stage of
the CLIC baseline, with a $-80\%$ polarised electron beam. The grey
area is excluded by the constraints applied to the ISS, the global fit
to EWPO and low-energy data~\cite{Fernandez-Martinez:2016lgt} being
the dominant constraint.

The ISS contribution vanishes, as expected, for a large seesaw scale
$M_R^{}$ and for vanishing Yukawa couplings. For a large fraction of
the parameter space, deviations of at least $-20\%$ are allowed, and
they reach a peak of $-38\%$, interestingly for Yukawa couplings
$|Y_\nu^{}|\sim 1$ and a seesaw scale of a few TeV. The ISS
deviation is then decreasing when approaching the region of excluded
points, reaching $\Delta^{\rm BSM}_{} = -25\%$. Using our previous
analysis of the dependence on the seesaw parameters, we have devised
the following approximate formulae to reproduce $\Delta^{\rm  BSM}$ in
the region allowed by the experimental constraints and for
$M_R>3$~TeV,
\begin{align}
  \mathcal{A}^{\rm ISS}_{\rm approx} =\,
  & \frac{(1~\text{TeV})_{}^2}{M_R^2} {\rm   Tr} (Y_\nu^{}
    Y_\nu^\dagger)\, \left(17.07 -
    \frac{19.79~\text{TeV}_{}^2}{M_R^2}\right),\nonumber\\
 \Delta^{\rm BSM}_{\rm approx} =\,
  & (\mathcal{A}^{\rm ISS}_{\rm approx})_{}^2 -11.94\, \mathcal{A}^{\rm
    ISS}_{\rm approx}.
 \label{eq:approx}
\end{align}
The coefficients (calculated here for $\sqrt{s} = 3$~TeV) depend on
the kinematics of the process and in particular on the c.m.~energy. We
have checked that, for $M_R^{} > 3$~TeV, our fit reproduces the full
result within $1\%$ in the region where the numerical error of our
calculation is negligible. For $M_R^{} < 3$~TeV, higher-order terms in
$1/M_R^{}$ that we have not included for simplicity and clarity give
sub-leading corrections that degrade the agreement between our fit and
the full calculation. For example, we find for our benchmark scenario
with $|Y_\nu^{}| = 1$ and $M_R^{} = 2.4$~TeV, the fit deviates from
the full result by $6\%$ only. Beside, below $M_R^{}<1.8$~TeV in the
allowed region, the fit is off the full results by around $\pm 10\%$
or more and we advise not to use it: We get for example for
$|Y_\nu^{}| = 0.7$ and $M_R^{}=1.8$~TeV the exact result $\Delta^{\rm
  BSM}_{} = -38.4\%$ to be compared to the result of our fit
$\Delta^{\rm BSM}_{\rm approx} = -34.8\%$, that is a $9\%$
difference. Compared to a similar map we derived
in~\cite{Baglio:2016bop} using the triple Higgs coupling, the coverage
with significant deviations is much larger.

\begin{figure*}[t]
  \centering
  \includegraphics[scale=0.71]{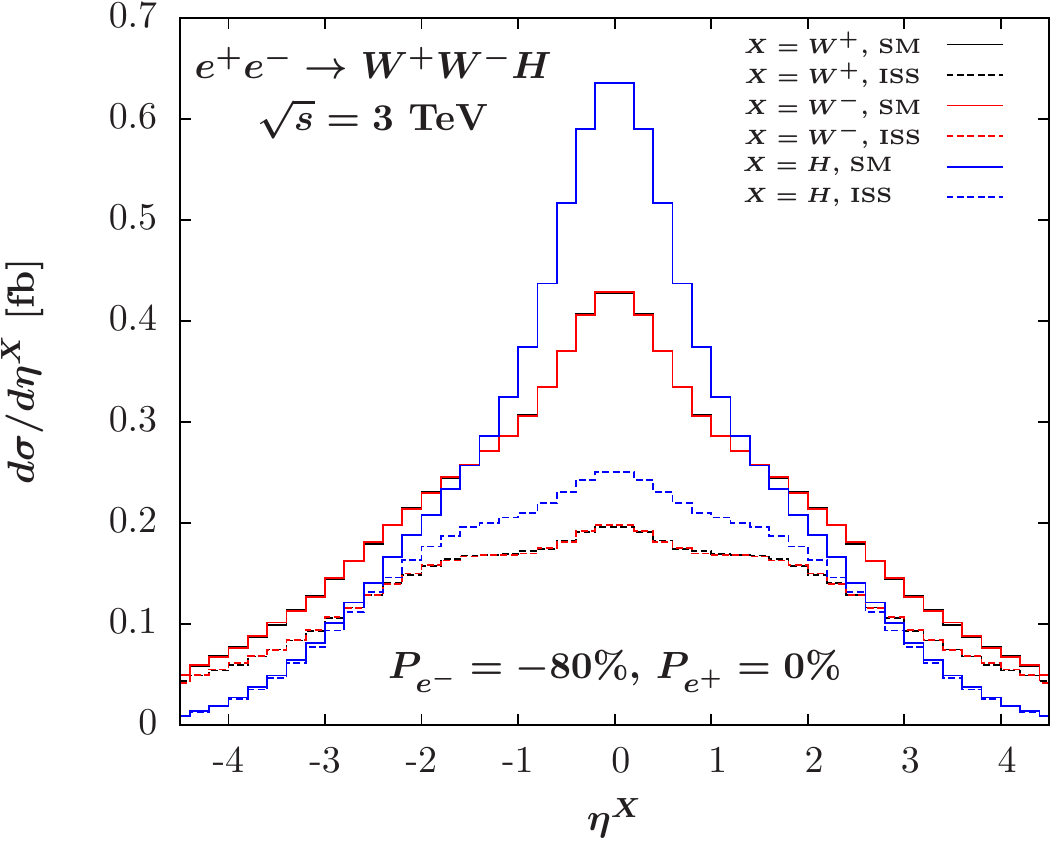}
  \hspace{3mm}
  \includegraphics[scale=0.71]{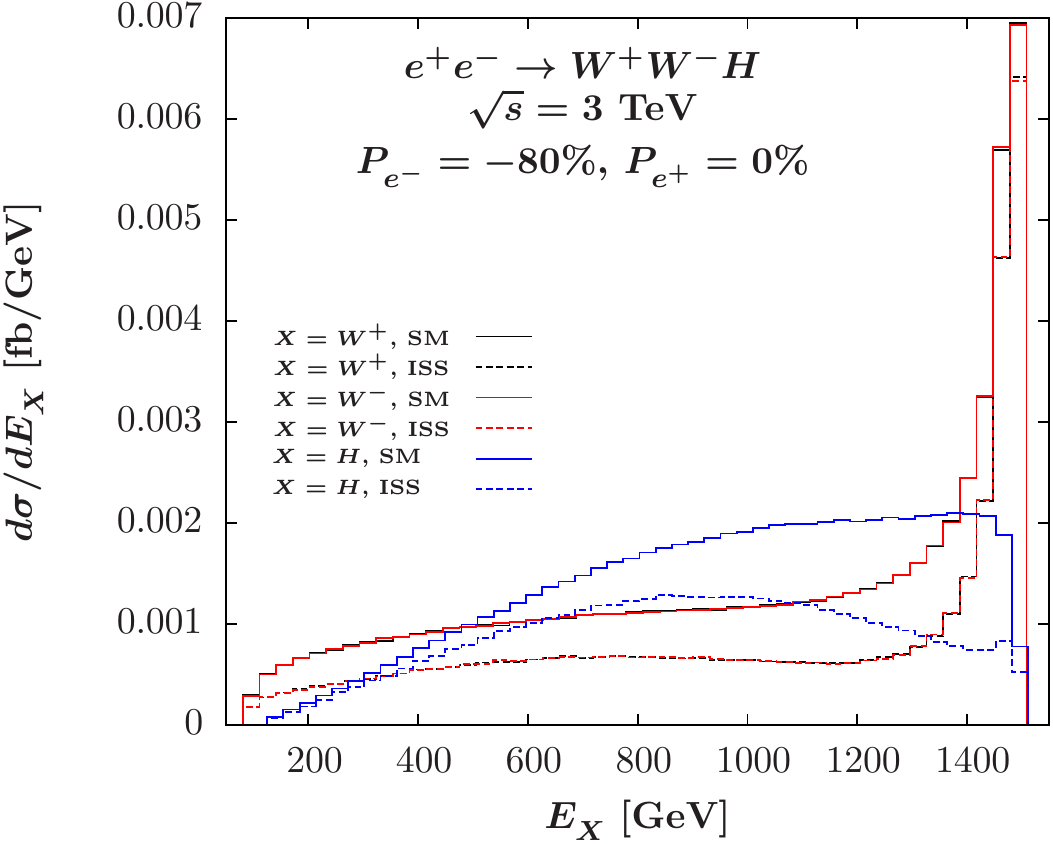}
  \caption[]{Pseudo-rapidity (left) and energy (right) distributions
    of the $W^+_{}$ (black), $W^-_{}$ (red) and Higgs (blue) bosons in
    the process $e^+_{}e^-_{}\to W^+_{}W^-_{}H$ at a c.m.~energy of 3
    TeV, using a $-80\%$ polarised electron beam. The solid curves
    stand for the SM predictions, the dashed curves stand for the ISS
    predictions using the benchmark scenario described in the text.}
  \label{fig:xsdists}
\end{figure*}

We have also considered kinematic distributions, in particular
the energy and pseudo-rapidity distributions of the final-state
particles. They are presented in fig.~\ref{fig:xsdists} in the
benchmark scenario we have already chosen for
fig.~\ref{fig:xsbenchmarkf1} using eq.(\ref{eq:hierarchical}). The
solid curves represent the SM predictions while the\linebreak dashed curves
stand for the ISS distributions. The $W^+_{}$ (in black) and $W^-_{}$
(in red) distributions are identical for both the pseudo-rapidity
(left) and the energy (right) observables, while the Higgs
distributions are displayed in blue.

For both $W^\pm_{}$ and Higgs boson, the pseudo-rapidities in the
central region have a different behaviour in the SM and in the
ISS. More specifically, the ISS corrections are substantial in the
region $|\eta|<1$. In the case of the energy spectrum, depicted on the right-hand
side of fig.~\ref{fig:xsdists}, the ISS correction is distributed over
the whole range for the $W^\pm_{}$ bosons, while it starts to be more
significant above 1~TeV for the Higgs boson. We have then considered the following
two cuts on the cross-section, in order to enhance $\Delta^{\rm
  BSM}_{}$:\linebreak $|\eta_{H/W^\pm_{}}^{}| < 1$ and $E_H^{} >
1$~TeV. Starting from polarised cross-sections $\sigma_{\rm pol}^{\rm
  SM} = 1.96$~fb and $\sigma_{\rm pol}^{\rm ISS} = 1.23$~fb, giving
$\Delta^{\rm BSM}_{} = -38\%$, we obtain the polarised cross-sections
$\sigma_{\rm pol,~cuts}^{\rm SM} = 0.42$~fb and $\sigma_{\rm
  pol,~cuts}^{\rm ISS} = 0.14$~fb, resulting in $\Delta^{\rm BSM}_{} =
-66\%$. This could potentially enlarge the region of interest for the
parameter space. We have also checked that the same type of
enhancement holds for another choice of the parameter point in the
region where $|Y_\nu^{}| \sim 4$ and $M_R^{} =
  8.3$~TeV. Using the same set of cuts we get a deviation of $-34\%$
instead of $-26\%$ for the cross-section without cuts. The level of
enhancement is reduced compared to the benchmark scenario with
$|Y_\nu^{}|=1$ because of the shape of the ISS $\eta$ distributions
which is closer to that of the SM prediction.

\section{Summary and outlook}

In this article we have investigated the effects of heavy neutrinos on the
production of a pair of $W$ bosons in association with a Higgs boson
at a lepton collider, $e^+_{}e^-_{}\to W^+_{}W^-_{} H$. After taking into
account the constraints on the model we have found sizeable deviations
that are maximal at a c.m.~energy of 3 TeV corresponding to the last
stage of the CLIC baseline, reaching a 38\% decrease of the
cross-section with respect to the SM prediction, in regions of the
parameter space with Yukawa couplings $|Y_\nu^{}|\sim 1$ and a seesaw
scale of a few TeV.
Analysing the kinematic distributions, we have found that
the negative deviations can be enhanced when using suitably chosen
cuts on the cross-section and reach $-66\%$. This is the
first time the effects of an extended neutrino sector on
 the
production cross-section of a pair of $W$ bosons in association with a
Higgs boson at a lepton collider have been investigated and our results
highlight the potential of this observable to beat future LHC
measurements which lose sensitivity in the high mass
regime~\cite{Cai:2017mow}.
They also demonstrate the ability of this
process to probe the coupling to the Higgs boson which is common to
all see-saw type I and III and their extensions, and motivate a
detailed sensitivity analysis of 
$e^+_{}e^-{}\to W^+_{}W^-_{} H$~\cite{inpreparation} as this could
provide a new, very competitive, and complementary observable to probe
neutrino mass models, especially in $\mathcal{O}(10)$~TeV mass regimes
with diagonal and real $Y_\nu^{}$ that are
difficult to probe otherwise.\bigskip

\begin{acknowledgements}
J.~B. acknowledges the support from the
Institutional Strategy of the University of T\"ubingen (DFG, ZUK 63),
the DFG Grant JA 1954/1, the Kepler Center of the University of
T\"ubingen, as well as the support from his Durham Senior Research
Fellowship COFUNDed between Durham University and the European Union
under grant agreement number 609412. S.P. and C.W. receive financial
support from the European Research Council under the European Union’s 
Seventh Framework Programme (FP/2007-2013)/ERC Grant NuMass Agreement
No. 617143. S.P. would also like to acknowledge partial support from
the European Union’s Horizon 2020 research and innovation programme
under the Marie Sk\l{}odowska-Curie grant agreements
No 690575 (RISE InvisiblesPlus) and No 674896 (ITN ELUSIVE),
from STFC and from the Wolfson Foundation and the Royal Society.
\end{acknowledgements}

{\small
\bibliographystyle{apsrev4-1}
\bibliography{eewwh_iss_paper}
}

\end{document}